# A similarity in patterns of global seismicity after St. Patrick's Day geomagnetic storms of 2013 and 2015

**Dimitar Ouzounov[1*], Galina Khachikyan[2]**

[1]Institute for Earth, Computing, Human and Observing, Chapman University, CA, USA , Dim.Ouzounov@gmail.com

[2]Institute of Seismology, Almaty, Kazakhstan, galina.khachikyan@gmail.com

\*       Correspondence: Dim.Ouzounov@gmail.com

**Abstract:** We present the results of a response of global seismicity to St. Patrick's Day (March 17) geomagnetic storms in 2013 and 2015, which occurred during rather similar solar flux levels and nearly identical storm sudden commencement times. A similar pattern of most substantial earthquake occurrence after storms is revealed. Namely, with a time delay of ~30 and ~39 days after storm onsets in 2013 and 2015, respectively, the strong crust earthquakes occurred at continental areas in Iran (M7.7, April 16, 2013) and Nepal (M7.8, April 25, 2015). Then, with a time delay of ~68 and ~74 days after storm onsets in 2013 and 2015, respectively, the strong deep-focused earthquakes occurred beneath the Sea of Okhotsk (M8.3, May 24, 2013, Russia) and beneath the Pacific Ocean (M7.8, May 30, 2015, Japan). It is shown that in the time of geomagnetic storm onsets (06:04 UT in 2013 and 04:48 UT in 2015), the longitudes at which the future strong earthquakes occurred were located under the polar cusps where the solar wind plasma would have direct access to the Earth's environment. The results support our earlier findings [Ouzounov and Khachikyan, 2022] that seismic activity may respond to geomagnetic storm onset with a time delay of up to ~ two months.

1. Introduction

Almost 170 years ago, Johann Rudolf Wolf- a Swiss astronomer and mathematician, claimed that sunspots could influence the occurrence of earthquakes [Wolf, 1853]. Currently, this idea is supported by many authors, for example, by Marchitelli et al. [2020], who showed a correlation between solar activity and large earthquakes worldwide. Nevertheless, overall, this topic, up to now, is considered controversial. So, Odintsov et al. [2006] reported that the number of earthquakes is highest during the solar-cycle sunspot maximum, but Simpson [1967] and Huzaimy and Yumoto [2011] declared that the seismicity is highest in the declining phase and minimum of the solar cycle**.**

A geomagnetic storm is one of Earth's most striking manifestations of solar activity. Nevertheless, the results of earthquakes triggered by geomagnetic storms are also considered controversial. So, Urata et al. [2018] investigated 4666 earthquakes with magnitudes M ≥ 6 that occurred during 1932-20    in the Pacific Rim region and concluded that external forces acting at the time of geomagnetic storms might trigger an earthquake. However, Yesugey [2009] investigated 122 838 earthquakes with magnitudes from 3.0 to 7.9 that occurred in 1965-2005 years in the region of the Anatolian peninsula and concluded that there is no finding that shows that the earthquakes occur as a result of a triggering under the effect of the geomagnetic storm. Recently, Akhoondzadeh and De Santis [2022] also applied the hypothesis that there is a correlation between solar-geomagnetic activities and powerful earthquakes by considering the variations of indices, including F10.7, Kp, ap, and Dst, before 333 large earthquakes (Mw ≥ 7.0) that occurred between 1 January 2000 and 28 April 2022. They found that anomalies in solar and magnetic indices were observed in 33% of earthquakes one day before the occurrence but analyzed 100 simulated data sets, and they found similar anomalies. They concluded that there is no significant correlation between solar and geomagnetic indices and the occurrence of strong earthquakes.

Love and Thomas [2013] have examined the claim that solar-terrestrial interaction, as measured by sunspots, solar wind velocity, and geomagnetic activity, might play a role in triggering earthquakes. They considered possible time lags between solar terrestrial variables and seismicity, which were equal to ±5 days for daily variations. As they concluded, they could not reject the null hypothesis of no solar-terrestrial triggering of earthquakes. It should be noted that Love and Thomas [2013] considered time lags of ±5 days, while several papers show that it may vary and sometimes may be longer.

So, Sobolev et al. [2001] investigated seismicity variations in the Northern Tien-Shan in 1975-1996 before and after geomagnetic storms with a sudden storm commencement (SSC). They showed that in the territories underlying the rocks with low electrical resistivity, the seismic activity increases, on average, 2 - 6 days after geomagnetic storm onset. Chen et al. [2020] investigated the correlation between geomagnetic storms (disturbed storm time index – Dst) and M ≥ 7.0 global earthquakes during 1957–2020; their results gave statistically significant evidence that the probability of geomagnetic storms increased around 26–27 days before earthquakes. Ouzounov and Khachikyan [2022] found that a time lag could be more than two months and more.

Considering the possibility of such long-time lags, it is difficult to reveal a seismicity response to a particular magnetic storm because recurrent storms, for example, may repeat with a time interval of ~27 days due to Sun's rotation. Moreover, the impact of the solar wind on Earth's environment may vary with time and season [Russell and McPherron, 1973]. Since geomagnetic storms occur in different seasons and start at different times, the response of seismic activity to other geomagnetic storms can be different. To solve this problem, analyzing the seismic response to geomagnetic storms occurring on the same days and at approximately the same time makes sense. Examples of such events include the St. Patrick's Day geomagnetic storms in 2013 and 2015 [Zhang et al., 2017]. Our work deals with a response of global seismicity to the St. Patrick's Day geomagnetic storms in 2013 and 2015.

**2. Data and Method**

The intensity of geomagnetic storms is often described with the 1-hour Dst (Disturbance Storm Time Index), which reflects the geomagnetic field variation averaged for several geomagnetic observatories. The same as Dst is the *SYM-H* index which has the distinct advantage of having a 1-min time resolution compared to the 1-hour time resolution of Dst. Both the Dst and SYM-H are indices designed to measure the intensity of the storm time ring current. As Wanliss and Showalter [2006] concluded, future studies use the SYM-H index as a de facto high-resolution Dst index. Therefore, this study used a 1-minute *SYM-H* index from World Data Center, Kyoto, available on the CDAWeb (http://cdaweb.gsfc.nasa.gov). The seismic data in this study include global M ≥ 4.5 earthquakes that occurred during six months in 2013 and 2015 after St. Patrick's Day (March 17) from the United States Geological Survey (USGS) catalog (https://earthquake.usgs.gov/earthquakes/search/). Each earthquake released seismic energy ($E$) was calculated using the $E=10^{(11.8+1.5M)}$, where $M$ is the earthquake's magnitude. Then, daily amounts of globally released seismic energy ($E_s$) were estimated for six months in 2013 and 2015, from March 17 to September 17. Temporal variations of daily released global seismic energy during investigated time intervals were analyzed, and similarity in their patterns was revealed.

**3. Results**

The St. Patrick's Day geomagnetic storms of 2013 and 2015 have attracted much community attention, and they have been the subject of several dedicated workshops and conference sessions [Zhang et al., 2017]. In particular, a Journal Geophysical Research unique collection (June 2017, Issue 6) includes 31 papers written by researchers worldwide, which are focused on characterizing and understanding storm-time geospace behavior during these specific events. These two events, especially the 2015 event being the strongest over the 24[th] solar cycle, are believed to be an excellent test bed for current theories and concepts [Zhang et al., 2017].

According to the LASCO (Large Angle and Spectrometric Coronagraph Experiment) on the board of SOHO (Solar and Heliospheric Observatory), the St. Patrick's Day geomagnetic storms on 17 March 2013 and 2015 originated from coronal mass ejections (CME) associated, respectively, with solar flare type C1.6 on March 15, 2013, at 07:12 U.T., and solar flare type C9 on March 15, 2015, at 02:36 U.T. (https://izw1.caltech.edu/ACE/ASC/DATA/level3/icmetable2.htm). Both CME was followed by the magnetic cloud [Wu et al., 2016; Clilverd et al., 2020].

It was shown by [Shreedevi et al., 2020] that the St. Patrick's Day (March 17) magnetic storms in 2013 and 2015 occurred in similar solar flux levels ($F_{10.7}$ = 118 and $F_{10.7}$ = 126, respectively) and were responsible for somewhat similar peak values of the Auroral Electrojet index (A.E.) equal to 2,700 nT and 2,200 nT, respectively. According to the OMNI database (http://cdaweb.gsfc.nasa.gov), a storm on March 17, 2013, started from a sudden storm commencement (SSC) at 06:04 U.T. with positive SYM/H index = +33 nT, the most significant negative SYM/H index = - 132 nT occurred at 20:28 U.T. on March 1 March 17, 2013, and duration of the main storm phase were equal to ~14 hours. A storm on March 17, 2015, started from SSC at 04:48 U.T. with a positive SYM/H index = 67 nT, the most significant negative SYM/H index = - 234 nT occurred at 22:47 U.T. on March 17, 2015, and the duration of the main storm phase was equal to ~18 hours. Figure 1 presents the 1-minute data of the SYM/H index from March 14 to March 26 in 2013 (black) and 2015 (red) from the OMNI database (http://cdaweb.gsfc.nasa.gov). It is seen that the storm of March 2015 was more substantial in comparison with the storm of 2013, which results evidently from different types of solar flares which generated these storms (C1.6 and C9, respectively). It was mentioned above that both storms occurred during similar solar flux levels and nearly identical storm sudden commencement times; they had compared the duration of the main storm phase and a somewhat similar shape and duration of a recovery phase, as seen in Figure 1.

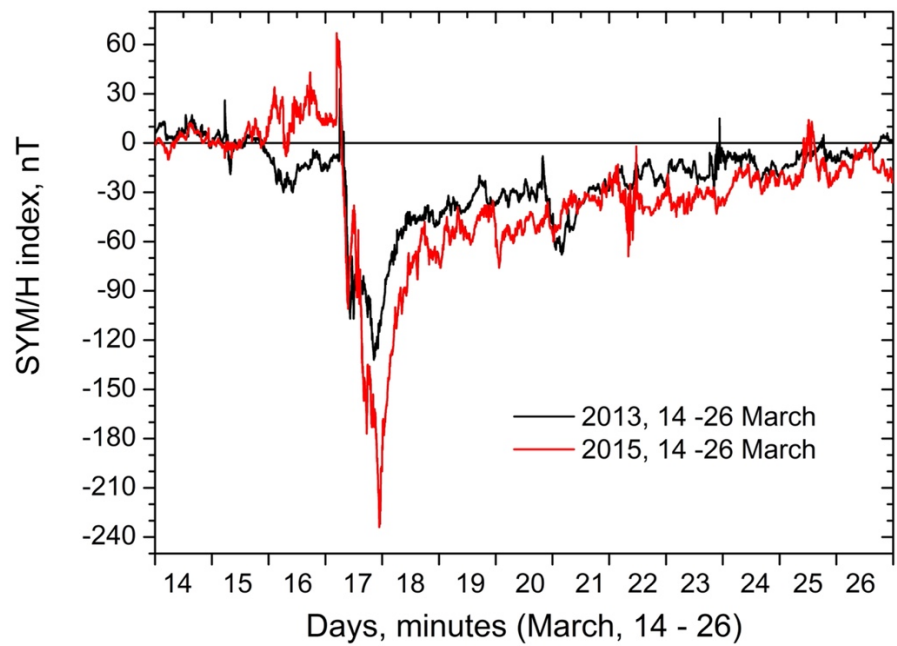

*Figure 1. The 1-minute SYM/H index for March 14 – 26 in 2013 (black) and 2015 (red) from the OMNI database(http://cdaweb.gsfc.nasa.gov).*

Observed similarities in the background conditions during the St. Patrick's Day geomagnetic storms on March 17, 2013, and March 17, 2015 occurrence, and the similarity in parameters of both storms (Figure 1) allows one to consider them as appropriate events to test the idea of solar-lithosphere relationships. Thus, we test a question: "If a response of global seismic activity to similar geomagnetic storms could be similar as well?" The result is presented below.

Figure 2 shows the histogram of the daily amount of globally released seismic energy (*Es*, in Joules) from March 17 to June 17 in 2013 and 2015 (black and red columns, respectively). Considered time intervals are restricted by June because the next strong geomagnetic storm occurred in June in 2013 (Dst= -137) and 2015 (Dst= -208).

Table 1 presents earthquake data, mainly responsible for forming Es-peaks marked as 1 and 2 in Figure 2. In Figure 2, dashed circles (1 and 2) mark the most significant peaks in the globally released seismic energy after March 17 in 2013 and 2015. For both years, the first peak occurred in April and the second in May.

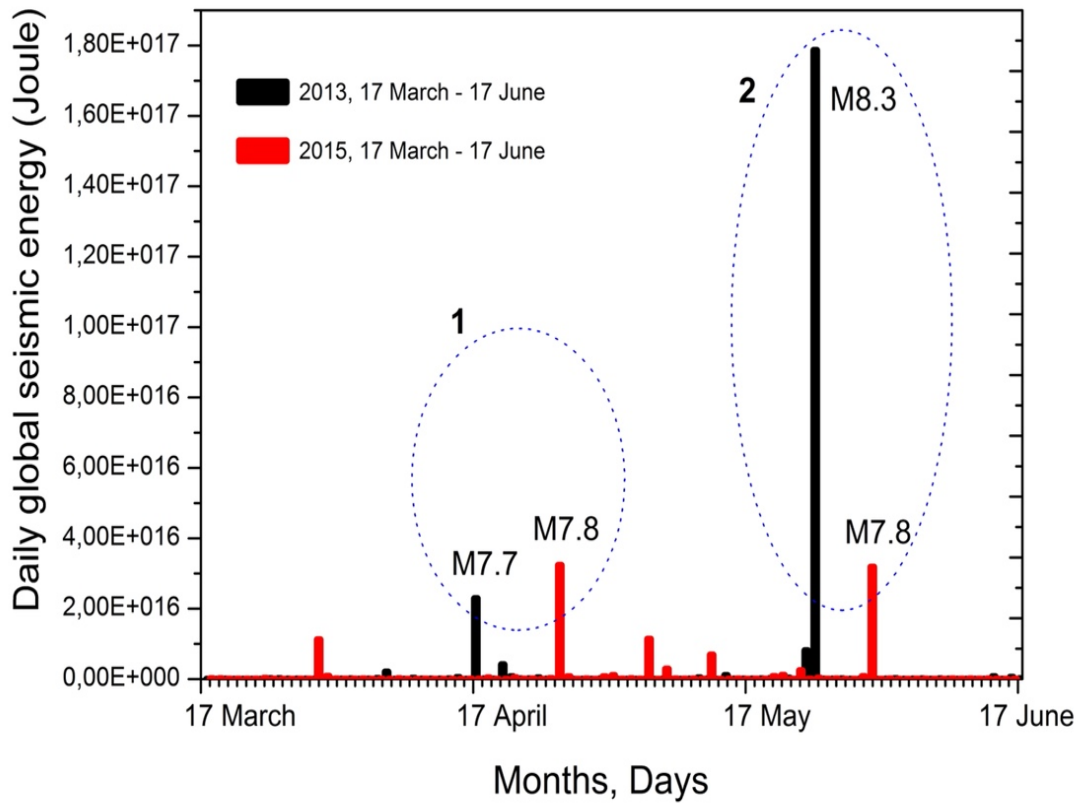

*Figure 2. The histogram of the daily amount of globally released seismic energy (Joules) from March 17 to June 17 in 2013 (black) and 2015 (red) were estimated using data on earthquakes with magnitude M ≥ 4.5 from the global USGS seismological catalog (https://earthquake.usgs.gov/earthquakes/search/).*

For *Es*-peak in April 2013, the M 7.7 earthquake is mainly responsible, which occurred on April 16, 2013, at an intermediate depth in the Arabia plate lithosphere, approximately 80 km beneath the surface in southeastern Iran. Here regional tectonics are dominated by the collisions of Arabia and Indian plates with Eurasia. Arabia plate lithosphere is subducted beneath the Eurasia plate at the Makran Coast of Pakistan and Iran [Hayes et al., 2016]. For this event, the time delay between the geomagnetic storm onset (March 17) and earthquake occurrence equals 30 days.

For *Es*-peak in April 2015, the M 7.8 earthquake is mainly responsible, which occurred on April 25, 2015, in Nepal. This earthquake occurred due to thrust faulting on or near the main thrust interface between the subducting India plate and the overriding Eurasia plate to the north [Hayes et al., 2016]. For this event, the time delay between the geomagnetic storm onset (March 17) and earthquake occurrence equals 39 days.

*Table 1. The data on the strongest earthquakes that occurred after the St. Patrick's Day geomagnetic storms in 2013 and 2015 are mainly responsible for Es-peaks marked in Figures 2 as 1 and 2.*

| Earthquake Date (yr.-mon.-day) | Time UTC | Geographic latitude | Geographic longitude | Depth (km) | Magnitude | The delay between storm onset and earthquake occurrence (days) |
|---|---|---|---|---|---|---|
| **The strongest earthquakes responsible for *Es*-peaks in April** | | | | | | |
| 2013-04-16 | 10:44:20 | 28.033°N | 61.996°E | 80 | 7.7 | 30 |
| 2015-04-25 | 06:11:25 | 28.231°N | 84.731°E | 8.2 | 7.8 | 39 |
| | | | | | | |
| **The strongest earthquakes responsible for *Es*-peaks in May** | | | | | | |
| 2013-05-24 | 05:44:48 | 54.892°N | 153.221°E | 598.1 | 8.3 | 68 |
| 2015-05-30 | 11:23:02 | 27.839°N | 140.493°E | 664.0 | 7.8 | 74 |

For *Es*-peak in May 2013, the deep-focus M 8.3 earthquake is mainly responsible, which occurred on May 24, 2013, in Russia at a depth of approximately 600 km beneath the Sea of Okhotsk. At the location of this earthquake, the Pacific and North American plates are converging, resulting in the subduction of the Pacific plate beneath Eurasia at the Kuril-Kamchatka Trench [Hayes et al., 2016]. The time delay between this event's geomagnetic storm onset and earthquake occurrence equals 68 days. For *Es*-peak in May 2015, again, the deep-focus M 7.8 earthquake is mainly responsible, which occurred on May 30, 2015, in Japan at a depth greater than 660 km beneath the Pacific Ocean. The earthquake is located within the interior of the Pacific plate that subducts beneath Japan, starting at the Izu Trench, several hundred kilometers to the east of the event [Hayes et al., 2016]. For this event, the time delay between geomagnetic storm onset and earthquake occurrence equals 74 days.

**3. Discussion and Conclusion**

The magnetosphere is a cavity around the Earth that is formed as the solar wind (the stream of particles and magnetic field ejected by the Sun) encounters an obstacle, the Earth's magnetic field [Sanchez et al., 1990]. Figure 3 presents a schematic drawing of the Earth's magnetosphere, adapted from http://english.igg.cas.cn/NC/RN/201504/t20150413_146311.html. In the polar magnetosphere, there are regions where the solar wind plasma would directly access the inner magnetosphere and upper atmosphere - these are the polar cusps [Willis, 1969].

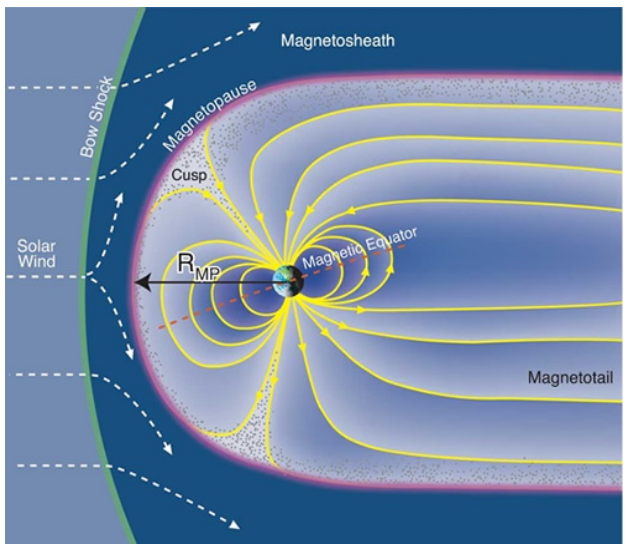

*Figure 3. A schematic drawing of the Earth's magnetosphere as adapted from http://english.igg.cas.cn/NC/RN/201504/t20150413_146311.html*

From a magnetic field point of view, the cusp is a funnel-shaped region where the high-latitude dayside (compressed) and night side (elongated) magnetic field lines converge toward the geomagnetic poles [Pitout & Bogdanova, 2021]. In schematic Figure 3, the yellow dots indicate these regions (northern and southern polar cups).

Investigations of the shocked solar wind penetration through the polar cusps is one of the primary science objectives of the Cluster mission, which is composed of four identical spacecraft flying in a tetrahedral formation around Earth (Escoubet et al., 2001). The Cluster data revealed the cusps as highly dynamic regions whose location is ruled by the interplanetary magnetic field (IMF) orientation, the solar wind pressure, and the reconnection line (X-line) length at the magnetopause. Figure 4, adapted from [Pitout et al., 2006], shows that most of the time, the cusp is located between 75° and 80° invariant magnetic latitudes and between 10h and 14h Magnetic Local Time (MLT), but sometimes its longitudinal extension may be more comprehensive depending, for example, on the length of the reconnection X-line [Crooker, 1979].

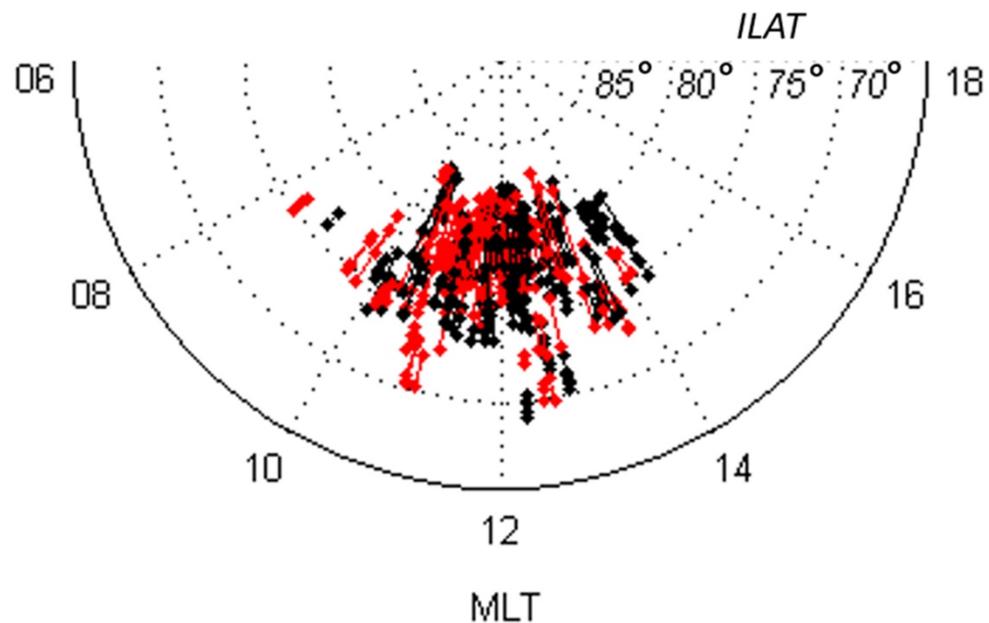

*Figure 4. Location of all cusp crossings by the Clusters as a function of magnetic local time (MLT) and invariant latitude (ILAT). Black and red designate crossings in the Northern and Southern Hemisphere, respectively (adapted from Pitout et al., 2006].*

On St. Patrick's Day, March 17, 2013, a geomagnetic storm started at 06:04 U.T. Using the online program (https://omniweb.gsfc.nasa.gov/vitmo/cgm.html), we calculated for UT = 06:04 the MLT values in points with coordinates (28.033°N, 61.996°E) and (54.892°N, 153.221°E) that are future epicenters of strong earthquakes in Iran and Okhotsk sea in 2013. On St. Patrick's Day, March 17, 2015, a geomagnetic storm started at 04:48 U.T. Using the same approach, we calculated for UT = 04:48 the MLT values in points with coordinates (28.231°N, 84.731°E) and (27.839°N, 140.493°E) that are future epicenters of strong earthquakes in Nepal and Japan in 2015.

It is obtained that in the time of geomagnetic storms onset on March 17 in, 2013, and 2015:
MLT=10.47 h at the area of the future epicenter in Iran in 2013;
MLT=15.65 h at the area of the future epicenter in Okhotsk Sea in 2013;
MLT=10.54 h at the area of the future epicenter in Nepal in 2015;
MLT=13.83 h at the area of the future epicenter in Japan in 2015.
These MLT values are presented in Table 2 (column 5). Taking into account the results of the Cluster mission (Figure 4), one may conclude that in the onset times of the St. Patrick's Day geomagnetic storms on 17 March 2013 and 2015, the longitudes at which the future strong earthquakes occurred in Iran, Okhotsk Sea, Nepal, and Japan were located under or near the polar cusp. Except for four considered strong earthquakes following the St. Patrick's Day geomagnetic storms on March 17, 2013, and 2015, there were ten other seismic events with M≥7.5 in 2013 and 2015. In particular, after the St. Patrick's Day geomagnetic storm on March 17, 2015, the two M7.5 earthquakes occurred in Papua New Guinea on March 29 and May 5, 2015, 12 and 49 days, respectively, respectively, later the storm's onset. The longitudes of their epicenters were located under or near the polar cusp at the time of magnetic storm onset (MLT=14.97 h and 14.93 h, respectively, Table 2).

*Table 2. Data on the geomagnetic storm, strong (M≥7.5) earthquake following this storm, the time delay between storm onset and earthquake occurrence, and magnetic local time (MLT) at the area of the future epicenter in the moment of geomagnetic storm onset.*

| # | Data on geomagnetic storm (date, time of onset (UT); maximal positive and negative values of the SYM/H- index (nT) | Data on strong (M≥7.5) earthquake following the geomagnetic storm (region, date, coordinates, depth-h km; magnitude) | The time delay between geomagnetic storm onset and earthquake occurrence (days) | Magnetic local time (MLT) in the area of the future epicenter in a moment of geomagnetic storm onset. |
|---|---|---|---|---|
| 1 | 2 | 3 | 4 | 5 |
| 1. | **March 17, 2013**; 06:04; SYM/H: +33; **-132.** | **Iran**, 2013-04-16; 10:44:20; 28.033°N, 61.996°E; h=80; **M7.7** | 30 | 10.47 |
| | | **Okhotsk Sea, Russia**, 2013-05-24; 05:44:48; 54.892°N, 153.221°E; h=598; **M8.3** | 68 | 15.65 |
| 2. | **March 17, 2015**, 04:48; SYM/H: +67; **-234.** | **Nepal**, 2015-04-25; 06:11:25; 28.231°N, 84.731°E; h=8.2; **M7.8** | 39 | 10.54 |
| | | **Japan**, 2015-05-30; 11:23:02; 27.839°N, 140.493°E; h=664; **M7.8** | 74 | 13.83 |
| | | **Papua New Guinea**, 2015-03-29; 23:48:31; 4.729°S, 152.562°E; h=41; **M7.5** | 12 | 14.97 |
| | | **Papua New Guinea**, 2015-05-05; 01:44:06; 5.462°S, 151.875°E; h=55; **M7.5** | 49 | 14.93 |
| 3. | **June 22, 2015**, 18:37; SYM/H: + 88; **-208.** | **Chile,** 2015-09-16; 22:54:32; 31.573°S, 71.674°W; h=22.4; **M8.3** | 86 | 13.96 |
| | | **Peru,** 2015-11-24; 22:45:38; 10.537°S, 70.944°W; h=606.2; **M7.6** | 155 | 13.87 |
| | | **Peru,** 2015-11-24; 22:50:54; 10.060°S, 71.018°W; h=620.6; **M7.6** | 155 | 13.86 |
| 4. | **October 31, 2012**; 16:52; SYM/H: +35; **-68.** | **Alaska**, 2013-01-05; 08:58:14; 55.228°N, 134.859°W; h=8.7; **M7.5** | 66 | 7.11 |
| 5. | **January 17, 2013**, 03:00; SYM/H: + 55; **- 57.** | **Solomon Islands,** 2013-02-06; 01:12:25; 10.799°S, 165.114°E; h=24; **M8.0** | 20 | 14.0 |
| 6. | **August 4, 2013**, 06:11; SYM/H: + 24; **- 56.** | **Pakistan**, 2013-09-24; 11:29:47; 26.951°N, 65.501°E; h=15; **M7.7** | 51 | 10.71 |
| 7. | **August 27, 2013**, 16:00; SYM/H: +13; **-68.** | **Scotia Sea**, 2013-11-17; 09:04:55; 60.274°S, 46.401°W; h=10; **M7.7** | 82 | 12.72 |
| 8 | **September 20, 2015**, 06:18; SYM/H: +3; **-84.** | **Afghanistan,** 2015-10-26; 09:09:42; 36.524°N, 70.368°E; h=231; **M7.5** | 36 | 11.21 |

On June 22, 2015, a strong magnetic storm (SYM/H = - 208 nT) occurred. Figure 5 shows for June 21-23, 2015, the 1-minute data on the magnitude of the averaged magnetic field vector on Earth's orbit, the vertical component of the Interplanetary Magnetic Field in the GSM coordinates (Bz), speed and density of the solar wind flux, pressure of the solar wind at the magnetopause, and geomagnetic SYM/H index from the OMNI database available on the CDAWeb (http://cdaweb.gsfc.nasa.gov).

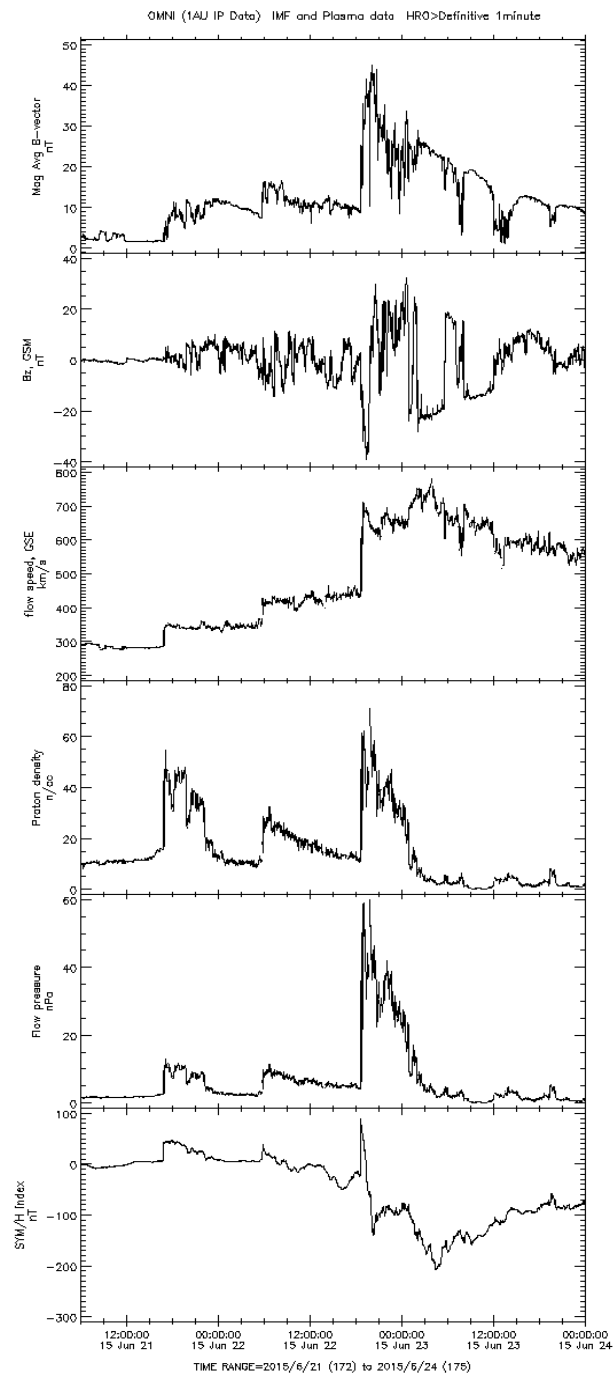

*Figure 5 (from top to bottom): The 1-minute data on the magnitude averaged magnetic field vector on Earth's orbit, the vertical component of the Interplanetary Magnetic Field (Bz) in the GSM coordinate system, speed and density of the solar wind flux, pressure of the solar wind at the magnetopause, and geomagnetic SYM/H index for June 21-23, 2015 from the OMNI database (http://cdaweb.gsfc.nasa.gov).*

It is evident from Figure 5 that a simultaneous strong jump-like growth in solar wind parameters occurred on June 22, 2015, at 18:37 UT; at this moment, the positive value of the SYM/H index reached up to 88 nT. At this time. the longitudes ~ 60°W-130°W were located under the polar cusp (MLT ~ 10h - 14 h). In analogy with the above-considered cases, one may expect that some weeks or months later, a strong earthquake may occur in this longitudinal region, and this is the case. On September 16, 2015, the M8.3 earthquake occurred in Chile with a time delay of ~86 days. At the location of the M8.3 Chile epicenter, the Nazca tectonic plate is moving towards the east-northeast concerning South America and begins its subduction beneath the continent at the Peru-Chile Trench [Hayes et al., 2016]. Two months later, on November 24, 2015, east of the Peru-Chile Trench, a "doublet" of strong (M7.6) deep-focused (h≥600 km) earthquakes occurred in Peru. At the time of storm onset (June 22, 2015, 18:37 UT), the longitudes of Chile and Peru earthquake epicenters were located under the polar cusp (MLT values were equal to 13.96 h, 13.87 h, and 13.86 h, respectively, Table 2).

For the rest five M≥7.5 events in 2013 and 2015, we also could identify the preceded geomagnetic storms when in times of storm onset, the longitudes of future epicenters were located under or near the polar cusp. So, the M7.5 earthquake in Alaska on January 5, 2013, was preceded ~66 days by a geomagnetic storm on October 31, 2012; the M8.0 earthquake in the Solomon Islands on February 6, 2013, was preceded ~20 days by a geomagnetic storm on January 17, 2013; the M7.7 earthquake in Pakistan on September 24, 2013, was preceded ~51 days by geomagnetic storm on August 4, 2013; the M7.7 earthquake at Scotia Sea on November 17, 2013, was preceded ~82 days by geomagnetic storm on August 27, 2013; and the M7.5 earthquake in Afghanistan on October 26, 2015, was preceded ~36 days by geomagnetic storm on September 20, 2015 (Table 2).

As seen in Table 2, powerful magnetic storms may be followed by many strong earthquakes. So, a storm on March 17, 2015 (SYM/H = -234 nT) was followed by four strong earthquakes; a storm on June 22, 2015 (SYM/H = -208 nT) was followed by three strong earthquakes; a storm on March 17, 2013 (SYM/H = - 132 nT) was followed by two strong earthquakes, while the five moderate to weak magnetic storms (SYM/H index ranges between -84 nT and -56 nT) were followed by only one M≥7.5 earthquakes.

Table 2 (column 5) shows that as a rule, at the time of geomagnetic storm onset (time of sharp compression of the magnetosphere by the shocked solar wind), the longitudes of the future epicenters are located under or near the polar cusp (10-14 MLT). The most significant disagreement occurred for two events: M8.3 in the Okhotsk Sea on May 24, 2013 (MLT=15.65) and M7.5 in Alaska on January 5, 2013 (MLT=7.11h). At the same time, it was reported by Crooker [1979] that the longitudinal cusp location may be some wider due to the much-extended length of the reconnection X-line at the magnetopause. Also, Pitout & Bogdanova [2021] noticed that the Clusters, which data are presented in Figure 4, due to the geometry of their orbits, do not allow to determine the entire longitudinal extent of the cusp correctly, and to do so, the spacecraft should have been placed in more or less parallel orbits, so that they fly at the same latitude but at different longitude or local times, that could be realized in the subsequent magnetosphere missions.

Column 4 in Table 2 shows that the delay between geomagnetic storm onset and following M≥7.5 earthquake varies between 12 and 155 days. This suggests that the shocked solar wind energy penetrates in some way into the lithosphere and modifies (prepares) the lithosphere to be realized in the form of an earthquake for several weeks or even months. It has been known for more than 30 years that the process of earthquake preparation modifies the state of the atmosphere and ionosphere, e.g. [Ouzounov et al. l, 2018, and references herein]. The time delay between the appearance of an anomaly in atmosphere-ionosphere parameters and earthquake occurrence may differ. For example, the multi-parameter satellite data analysis revealed anomalous atmospheric and ionospheric effects over the earthquake preparation zone of the Nepal M7.8 earthquake 21 days before April 25, 2015, main shock [Ouzounov et al., 2021]. It was noticed by [Ouzounov and Khachikyan, 2022] that with a time lag of about two months, the seismic activity increases near the footprint of the geomagnetic field line, belonging to a different radiation belt created in the lower magnetosphere by the precipitated high energy electrons from the outer radiation belt due to geomagnetic storm. Marchetti et al. [2022] analyzed the magnetic field and electron density data from the Swarm three-identical satellite constellation concerning global M5.5+ shallow earthquakes from November 2013 to November 2021. They have found that the anticipation time of the anomaly in the magnetic field and electron density increases with the magnitude of the incoming earthquakes following the Rikitake laws [Rikitake, 1987]. Marchetti et al. [2022] have revealed that the anticipation time of large earthquakes (M7.5+) may reach up to some years before an event occurrence. Thus, the 155 days of lithosphere preparation after a magnetic storm on June 22, 2015, before a "doublet" of M7.6 deep-focused earthquakes occurred in Peru on November 24, 2015, does not look as unexpected.

In Figure 6, we show a spatial distribution of 14 epicenters of strong (M≥7.5) earthquakes that occurred around the globe in 2013 and 2015 (stars). Blue quadrants, red circles, and purple triangles with digits 1, 2, and 3, respectively, indicate the location of geomagnetic poles in the Northern and Southern hemispheres for the magnetic epochs of 1975–2015 in three options: 1 – not displaced dipole (79.5° N, 71.4° W, and 79.5° S, 108.6° E); 2 – displaced dipole (80.3° N, 114.9° W, and 64.9° S, 138.4° E), 3 – corrected geomagnetic poles (81.46° N, 82.33° W and 74.18° S, 126.19° E) as calculated by the on-line program (Corrected Geomagnetic Coordinates and IGRF/DGRF Model Parameters https://omniweb.gsfc.nasa.gov/vitmo/cgm.html).

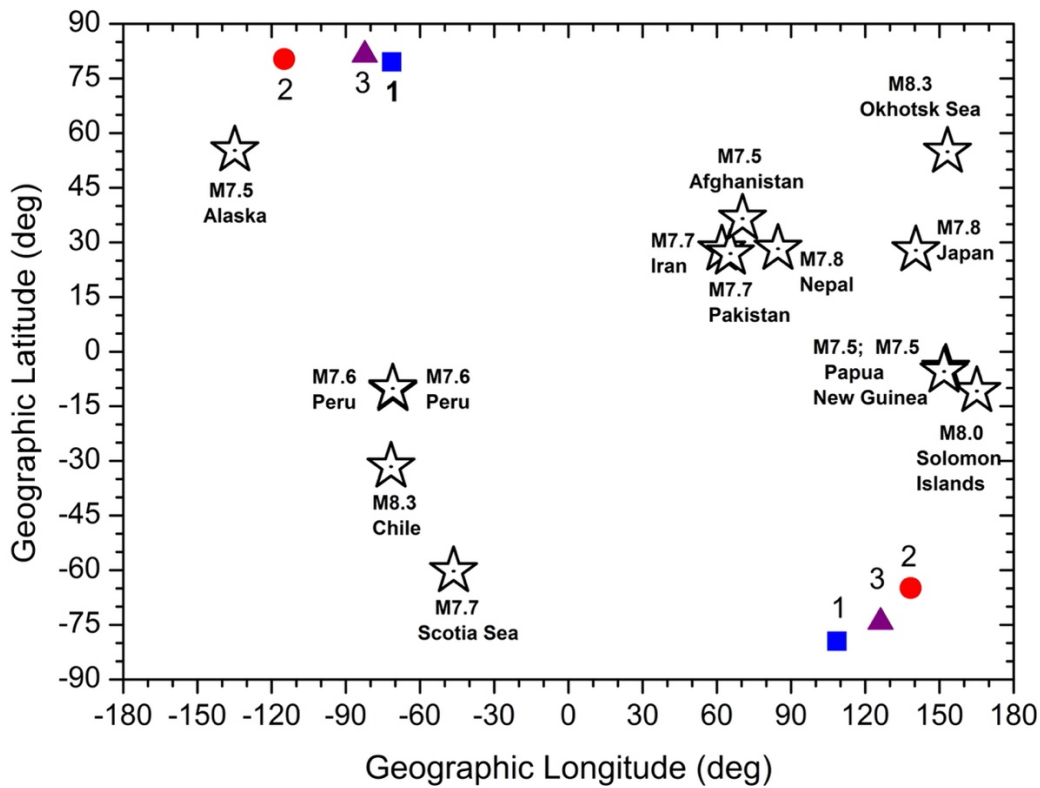

*Figure 6. Spatial distribution of epicenters of 14 strong (M≥7.5) earthquakes that occurred around the globe in 2013 and 2015 (stars); blue quadrants, red circles, and purple triangles with digits 1, 2, and 3, respectively, indicate the location of geomagnetic poles in the Northern and Southern hemispheres for three options: 1 – not displaced geomagnetic dipole, 2 – displaced geomagnetic dipole, 3 - corrected geomagnetic poles, see text.*

It can be seen from Figure 6 that the longitudinal regions where the strongest earthquakes occurred in 2013 and 2015 (America and East Asia) are closely confined to the longitudes of the location of the geomagnetic poles. Earthquake epicenters in the American continent are centered around longitudes of the location of geomagnetic poles in the Northern Hemisphere. In contrast, earthquake epicenters in East Asia are centered around longitudes of locations of geomagnetic poles in the Southern Hemisphere (color quadrants, circles, and triangles in Figure 6). It has long been well known that the strongest earthquakes on the planet occur precisely in these longitudinal regions (America and East Asia).

Figure 7 presents the histogram of the distribution of the number of strong (M≥7.5) earthquakes that occurred around the globe from 1973-2017 years (212 events) in dependence on longitude. Blue quadrants, red circles, and purple triangles mark the longitudes of the geomagnetic pole locations in the Northern and Southern hemispheres, respectively, as it was discussed above.

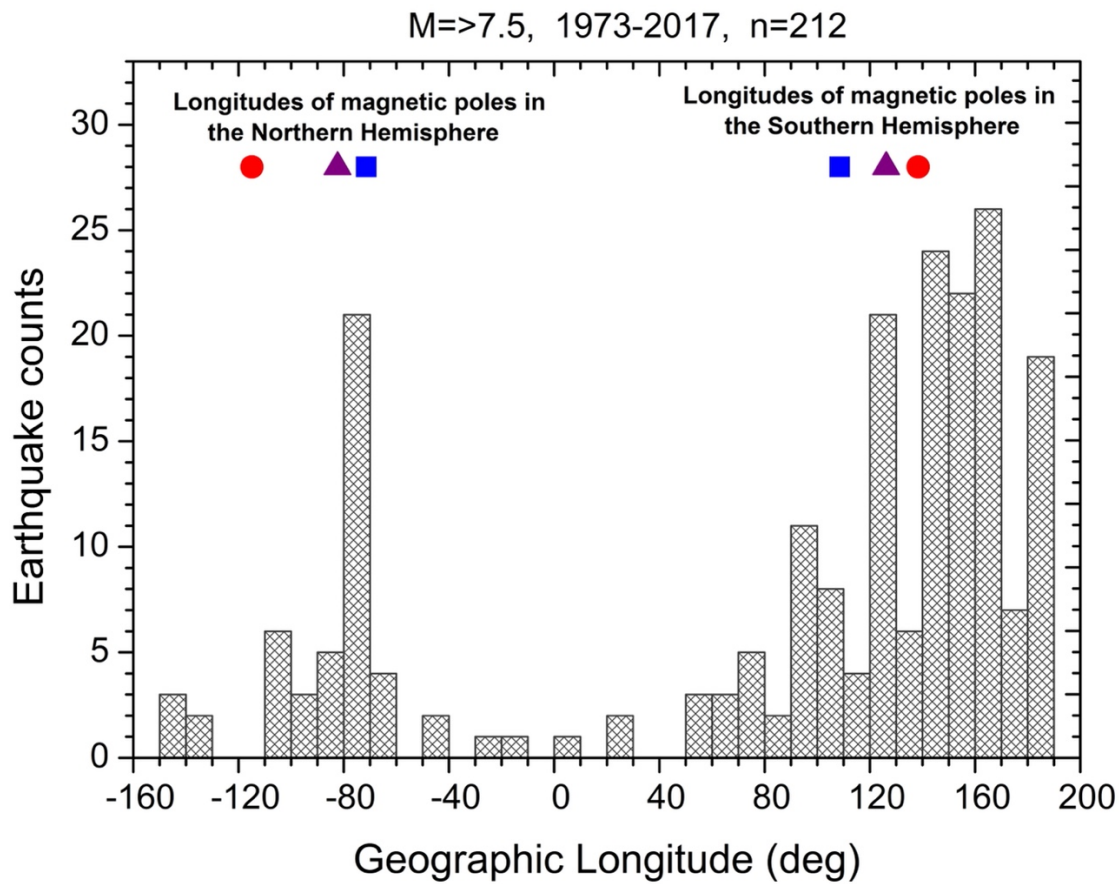

*Figure 7.* The histogram of the distribution of the number of strong (M≥7.5) earthquakes that occurred around the globe in 1973-2017 years (212 events) in dependence on longitude; blue quadrants, red circles, and purple triangles mark the longitudes of the geomagnetic pole's locations in the Northern and Southern hemispheres for three options, as it was indicated in Figure 6.

It can be seen from Figure 7 that at the globe, the longitudinal regions of the strongest earthquake occurrence are centered around the longitudes of the geomagnetic pole locations. The reason for this may be the polar cusps (Figure 4). It is expected, evidently, that the most significant amount of the shocked solar wind energy will penetrate the polar cusps in those times when the longitudes of the geomagnetic pole location will be strictly under the cusp.

**In conclusion**: There has yet to be an advanced idea of how the shocked solar wind energy can reach the lithosphere and what it does there for some weeks or even months before it will be realized in the kind of earthquake. Obviously, this issue should become a key one in solving the problems of solar-terrestrial physics.

**Author Contributions:** D.O. and G.K. provided the concepts of the manuscript. G.K. organized and wrote the manuscript. All authors provided critical feedback and helped shape the research, analysis, and manuscript. All authors have read and agreed to the published version of the manuscript.
**Funding:** This research received no external funding.
**Data Availability Statement:** Generated Statement: The original contributions presented in the study are included in the article/supplementary material, further inquiries can be directed to the corresponding author/s.
**Acknowledgments:** Thanks to the U.S. Geological Survey and European-Mediterranean Seismological Centre for providing earthquake information services and data. We acknowledge the use of NASA/GSFC's Space Physics Data Facility's CDAWeb service and OMNI data.

**Conflicts of Interest:** The authors declare that the research was conducted without any commercial or financial relationships that could be construed as a potential conflict of interest.

**References**

1. Akhoondzadeh, M.; De Santis, A. Is the Apparent Correlation between Solar-Geomagnetic Activity and the Occurrence of Powerful Earthquakes a Casual Artifact? Atmosphere 2022, 13, 1131. https:// doi.org/10.3390/atmos13071131
2. Chen H., Wang R., Miao M., Liu X., Ma Y., Hattori K., Han P. A Statistical Study of the Correlation between Geomagnetic Storms and M ≥ 7.0 Global Earthquakes during 1957–2020. *Entropy* 2020, *22*(11), 1270; https://doi.org/10.3390/e22111270
3. Clilverd, M. A., Rodger, C. J., van de Kamp, M., &Verronen, P. T. (2020). Electron precipitation from the outer radiation belt during the St. Patrick's day storm 2015: Observations, modeling, and validation. Journal of Geophysical Research: Space Physics, 125, e2019JA027725. https://doi.org/ 10.1029/2019JA027725
4. Clilverd, N. U. (1979). Dayside merging and cusp geometry. Journal of Geophysical Research, 84(A3), 951–959. https://doi.org/10.1029/ JA084iA03p00951
5. Escoubet, C. P., Fehringer, M., & Goldstein, M. (2001). The Cluster mission. Annals of Geophysics, 19, 1197–1200. https://doi.org/10.5194/ angeo-19-1197-2001.
6. Hayes, G.P., Myers, E.K., Dewey, J.W., Briggs, R.W., Earle, P.S., Benz, H.M., Smoczyk, G.M., Flamme, H.E., Barnhart, W.D., Gold, R.D., and Furlong, K.P., 2017, Tectonic summaries of magnitude 7 and greater earthquakes from 2000 to 2015: U.S. Geological Survey Open-File Report 2016–1192, 148 p., https://doi.org/10.3133/ofr20161192.
7. Huzaimy, J.M.; Yumoto, K. Possible correlation between solar activity and global seismicity. In Proceedings of the 2011 IEEE International Conference on Space Science and Communication, Penang, Malaysia, 12–13 July 2011; pp. 138–141.
8. Love, J.J.; Thomas, J.N. Insignificant solar-terrestrial triggering of earthquakes. Geophys. Res. Lett. 2013, 40, 1165–1170.
9. Marchetti, D.; De Santis, A.; Campuzano, S.A.; Zhu, K.; Soldani, M.; D'Arcangelo, S.; Orlando, M.; Wang, T.; Cianchini, G.; Di
10. Marchitelli V, Troise C, Harabaglia P, Valenzano B and De Natale G (2020) On the Long Range Clustering of Global Seismicity and its Correlation With Solar Activity: A New Perspective for Earthquake Forecasting. Front. Earth Sci. 8:595209. doi: 10.3389/feart.2020.595209
11. Mauro, D.; et al. Worldwide Statistical Correlation of Eight Years of Swarm Satellite Data with M5.5+ Earthquakes: New Hints about the Preseismic Phenomena from Space. Remote Sens. 2022, 14, 2649. https://doi.org/10.3390/rs14112649
12. Odintsov, S.; Boyarchuk, K.; Georgieva, K.; Kirov, B.; Atanasov, D. Long-period trends in global seismic and geomagnetic activity and their relation to solar activity. Phys. Chem. Earth 2006, 31, 88–93
13. Ouzounov, D., S. Pulinets, K.,Hattori, P.,Taylor. (Ed's) Pre-Earthquake Processes: A Multi-disciplinary Approach to Earthquake Prediction Studies, American Geophysical Union. Published by John Wiley & Sons, 2018, 385 p.
14. Ouzounov D, Pulinets S, Davidenko D, Rozhnoi A, Solovieva M, Fedun V, Dwivedi B.N., Rybin A, Kafatos M. and Taylor P. (2021) Transient Effects in Atmosphere and Ionosphere Preceding the 2015 M7.8 and M7.3 Gorkha–Nepal Earthquakes. Front. Earth Sci. 9:757358.doi: 10.3389/feart.2021.75735
15. Ouzounov D., G. Khachikyan. On the impact of the geospace environment on solar-lithosphere coupling and earthquake occurrence. 2022. arXiv:2202.08052v2 **[physics.geo-ph].** https://arxiv.org/abs/2202.08052
16. Pitout, F., & Bogdanova, Y. V. (2021). The polar cusp was seen by Cluster. Journal of Geophysical Research: Space Physics, 126, e2021JA029582. https://doi. org/10.1029/2021JA029582
17. Pulinets S., D. Ouzounov, A.Karelin, D.Boyarchuk Earthquake precursors in atmosphere and ionosphere. A new concept. Springer-Nature, 2022, 314pp
18. Rikitake, T. Earthquake precursors in Japan: Precursor time and detectability. Tectonophysics 1987, 136, 265–282.

19. Russell C. T. and R. L. McPherron, "Semi-Annual Variation of Geomagnetic Activity," J. Geophys. Res. 78, 92–108 (1973
20. Sanchez E. R., C.-I. Meng, and P. T. Newell. OBSERVATIONS OF SOLAR WIND PENETRATION INTO THE EARTH'S MAGNETOSPHERE: THE PLASMA MANTLE Johns Hopkins A PL Technical Digest, Volume 11 , Numbers 3 and 4 (1990) https://www.jhuapl.edu/content/techdigest/pdf/V11-N3-4/11-03-Sanchez.pdf
21. Shreedevi, P. R., Choudhary, R. K., Thampi, S. V., Yadav, S., Pant, T. K., & Yu, Y., et al. (2020). Geomagnetic storm-induced plasma density enhancements in the southern polar ionospheric region: A comparative study using St. Patrick's Day storms of 2013 and 2015. Space Weather, 18, e2019SW002383. https://doi.org/ 10.1029/2019SW002383
22. Simpson, J.F. Solar activity as a triggering mechanism for earthquakes. Earth Planet. Sci. Lett. 1967, 3, 417–425.
23. Sobolev, G. A., N. A. Zakrzhevskaya, and E. P. Kharin. On the relation between seismicity and magnetic storms. *Phys. Solid Earth*. 2001, *Volume* 37. 917–927.
24. Urata, N.; Duma, G.; Freund, F. Geomagnetic Kp Index and Earthquakes. Open J. Earthq. Res. 2018, 7, 39–52.
25. Wanliss, J. A., and K. M. Showalter (2006), High-resolution global storm index: Dst versus SYM-H, J. Geophys. Res., 111,A02202, doi:10.1029/2005JA011034
26. Willis, D. M. (1969). The influx of charged particles at the magnetic cusps on the boundary of the magnetosphere. Planetary and Space Science, 17(3), 339–348. https://doi.org/10.1016/0032-0633(69)90067-1
27. Wolf, R. On the periodic return of the minimum of sunspots: The agreement between those periods and the variations of magnetic declination. Philos. Mag. 1853, 5, 67.
28. Wu, C-C., Liou, K., Lepping, R.P., Hutting, L., Plunkett, S., Howard, R. A., & Socker, D. (2016). The first super geomagnetic storm of solar cycle 24: "The St. Patrick's Day eventMarch 17ch 2015)", Earth, Planets and Space, 68, 1.
29. Yesugey, S.C. Comparative Evaluation Of The Influencing Effects Of Geomagnetic Solar Storms On Earthquakes In Anatolian Peninsula. Earth Sci. Res. J. 2009, 13, 82–89.
30. Zhang, S.-R., Y. Zhang, W. Wang, and O. P. Verkhoglyadova (2017), Geospace system responses to the St. Patrick's Day storms in 2013 and 2015, J. Geophys.Res. Space Physics, 122, 6901–6906, doi:10.1002/2017JA024232.